\documentclass[%
 aip,
 sd,%
 amsmath,amssymb,
 reprint,%
]{revtex4-1}

\usepackage{color}
\usepackage{graphicx}
\usepackage{dcolumn}
\usepackage{bm}
\usepackage{verbatim}
\usepackage{tabularx}
\usepackage{amsmath}
\usepackage{float}
\usepackage{siunitx}
\usepackage{array,booktabs}

\begin{document}

\preprint{AIP/123-QED}

\title[Observation of Shock-Front Separation in Multi-Ion-Species Collisional Plasma Shocks]{Observation of Shock-Front Separation in Multi-Ion-Species Collisional Plasma Shocks}

\author{Tom Byvank}
\email[]{tbyvank@lanl.gov}
\affiliation{Los Alamos National Laboratory, Los Alamos, NM 87545}

\author{Samuel J. Langendorf}
\email[]{samuel.langendorf@lanl.gov}
\affiliation{Los Alamos National Laboratory, Los Alamos, NM 87545}

\author{Carsten Thoma}
\affiliation{Voss Scientific, Albuquerque, NM 87108}

\author{Scott C. Hsu}
\affiliation{Los Alamos National Laboratory, Los Alamos, NM 87545}

\date{\today}

\begin{abstract}
We observe shock-front separation and species-dependent shock widths in multi-ion-species
collisional plasma shocks, which are produced by 
obliquely merging plasma jets of a He/Ar mixture (97\% He and 3\% Ar by initial number density)
on the Plasma Liner Experiment [S. C. Hsu et al., IEEE Trans.\ Plasma Sci.~{\bf 46}, 1951 (2018)].
Visible plasma
emission near the He-I 587.6-nm and Ar-II 476.5--514.5-nm lines are simultaneously recorded by splitting
a single visible image of the shock into two different fast-framing cameras with
different narrow bandpass filters
($589 \pm 5$~nm for observing the He-I line and $500 \pm 25$~nm for the Ar-II lines). 
For conditions in these experiments (pre-shock ion and electron densities $\approx 5\times 10^{14}$~cm$^{-3}$,
ion and electron temperatures of $\approx 2.2$~eV, and relative plasma-merging speed of 
22~km/s), the observationally inferred magnitude of
He/Ar shock-front separation and the shock widths themselves are $< 1$~cm, which correspond
to $\sim 50$ post-shock thermal ion--ion mean free paths.
These experimental lengths scales are in reasonable qualitative and quantitative agreement with results from
1D multi-fluid simulations using the \textsc{chicago} code. However, there are differences between the experimentally-inferred and simulation-predicted ionization states and line emission intensities, particularly in the post-shock region.
Overall, the experimental and simulation results are consistent with theoretical predictions that 
the lighter He ions diffuse farther ahead within the overall shock front than the heavier Ar ions.

%
\end{abstract}

\maketitle

\section{\label{sec:intro}Introduction}

Supersonic flows generate shocks in astrophysics, aerodynamics, and high-energy-density (HED) plasma
experiments. Compared to hydrodynamic shocks in neutral gases, collisional plasma shocks contain ion and 
electron species, arise due to Coulomb collisions, and are influenced by electromagnetic fields 
\cite{Cowling1942, Jukes1957, Jaffrin1964, Center1967}. 
A plasma shock front with multiple ion species contains additional structure compared to a single ion plasma 
shock. Prior experiments, simulations, and theoretical work explored multi-ion-species effects in the 
context of inertial confinement fusion (ICF), for which species separation in the fusion fuel potentially
leads to neutron yield degradation \cite{Rygg2006, Herrmann2009, Casey2012, Rinderknecht2015, 
	Higginson2019, Sio2019, Larroche2012, Bellei2013, Bellei2014, Inglebert2014, 
	Vold2019, Amendt2010, Amendt2011, Kagan2014}. Interspecies ion separation and 
velocity separation have been experimentally observed \cite{Hsu2016, Joshi2017, Rinderknecht2018, 
	Rinderknecht2018a,Joshi2019}. Additional simulation and theoretical research on multi-ion-species plasmas 
examined how ion species diffusion causes species separation \cite{Kagan2012, Bellei2014a, Simakov2016, Glazyrin2016, Simakov2017, Keenan2017, Keenan2018}. The present research reports direct observations
of the spatial profile of a multi-ion-species shock, showing shock-front separation and
species-dependent shock widths in collisional plasma shocks. The experimental length scale results of the shock structure agree with 1D 
multi-fluid simulations using the \textsc{chicago} code \cite{Thoma2011, 
	Thoma2013, Thoma2017}. However, particularly in the post-shock region, there are differences between the experimentally-inferred and simulation-predicted ionization states and line emission intensities. These experimental and simulation results are both consistent 
with ion species diffusion theory.
This experimental data can be used to validate and benchmark numerical simulations of 
plasma environments with multi-ion-species collisional plasma shocks, especially
in HED, magneto-inertial-fusion (MIF), and ICF experiments.

The organization of this paper is as follows: Sec.~\ref{sec:background} provides background on the ion diffusion theory to which we compare our results, Sec.~\ref{sec:setup} describes the experimental setup of colliding plasma jets to generate a shock, Sec.~\ref{sec:exp} presents results from the experimentally observed shock profiles, Sec.~\ref{sec:lengths} overviews the relevant length scales in our experiments, Sec.~\ref{sec:sim} discusses the agreement between the multi-fluid simulations and the experiments, Sec.~\ref{sec:diff} gives estimates for the contributions from various diffusion mechanisms to describe our observations, and Sec.~\ref{sec:concl} highlights our conclusions.

\section{\label{sec:background}Background on Ion Diffusion Theory}

Theory and simulations about interspecies ion diffusion predict
that lighter ions diffuse farther ahead within a collisional plasma shock (closer to the pre-shock region) than 
heavier ions \cite{Larroche2012, Bellei2013, Inglebert2014, Vold2019, Amendt2010, Amendt2011, Kagan2014, Kagan2012, Bellei2014a, Simakov2016, Glazyrin2016, Simakov2017, Keenan2018}.
Following \textcite{Kagan2012}, in the center-of-mass frame of a multi-ion-species fluid element, the 
diffusion flux $\bar{f}_{1}$ of the lighter species 1 (in our case, He) equals the negative flux $\bar{f}_{2}$ of the 
heavier species 2 (in our case, Ar), 
\begin{equation}
\bar{f}_{1} = \rho_{1}\bar{v}_{D1} = -\bar{f}_{2} = -\rho_{2}\bar{v}_{D2},
\label{eqn:flux}
\end{equation} 
with mass densities $\rho_{1}$, $\rho_{2}$, and diffusion velocities $\bar{v}_{D1}$, $\bar{v}_{D2}$.
The diffusion flux is related to the species mass concentration time evolution (continuity equation), 
\begin{equation}
\rho \frac{\partial c_{m1}}{\partial t} + \rho \bar{u} \cdot \nabla c_{m1} + \nabla \cdot \bar{f}_{1} = 0,
\label{eqn:conc}
\end{equation}
with total mass density $\rho$, species 1 mass concentration $c_{m1} = \rho_{1}/ \rho$, 
and bulk fluid velocity $\bar{u}$.
The diffusion flux of the lighter species can be written as
\begin{equation}
\begin{split}
\bar{f}_{1} = &- \rho D \nabla c_{m1} - \rho D \left( \frac{\kappa_{p}}{P_{i}} \nabla P_{i} \right) \\ &- \rho D 
\left( \frac{\kappa_{Ti}}{T_{i}} \nabla T_{i} + \frac{\kappa_{Te}}{T_{e}} \nabla T_{e} \right) \\ &- \rho D \left( \frac{e 
	\kappa_{e}}{T_{i}} \nabla \Phi \right),
\end{split}
\label{eqn:diffusion}
\end{equation}
where the first term is the classical diffusion flux based on the mass concentration gradient, the second term is 
the barodiffusion flux based on the ion pressure gradient, the third term is the thermodiffusion flux based on 
the ion and electron temperature gradients, and the fourth term is the electrodiffusion flux based on the electric 
field (negative gradient of the electric potential). The new variables in Eq.~(\ref{eqn:diffusion}) are the classical diffusion 
coefficient $D$, barodiffusion ratio $\kappa_{p}$, total ion pressure $P_{i}$, ion thermodiffusion ratio
$\kappa_{Ti}$, electron thermodiffusion ratio $\kappa_{Te}$, ion temperature $T_{i}$ (approximating as equal for each species), electron temperature $T_{e}$, electrodiffusion ratio $\kappa_{e}$, and electric 
potential $\Phi$.

Within a shock front, the gradients of pressure, temperature, and electric potential point in the direction from the 
pre-shock region toward the post-shock region \cite{Jaffrin1964}. Additionally, for the He/Ar mixture in our experimental shocks, the various diffusion fluxes have the same sign except for the relatively
small electron thermodiffusion, as described in Sec.~\ref{sec:diff}. We also assume no initial concentration separation. Therefore, in the center of mass frame, 
Eqs.~(\ref{eqn:flux}) and (\ref{eqn:diffusion}) predict that the lighter species diffusion velocity $\bar{v}_{D1}$ 
points in the direction from the post-shock region toward the pre-shock region, and the heavier species 
diffusion velocity $\bar{v}_{D2}$ points in the opposite direction. In the present research, we directly observe 
the spatial profile of a plasma shock front containing a He/Ar mixture, offering the opportunity
to validate models of multi-ion-species shock evolution based on the theory.

\section{\label{sec:setup}Experimental Setup}

In order to experimentally observe shock-front separation in multi-ion-species collision plasma shocks, we form the shocks by obliquely merging two supersonic plasma jets, and we image the different species within the shock profiles using distinct narrow bandpass filters. In this section, we describe this experimental setup.

\begin{figure}[!tb]
	\includegraphics[width=3.25truein,keepaspectratio]{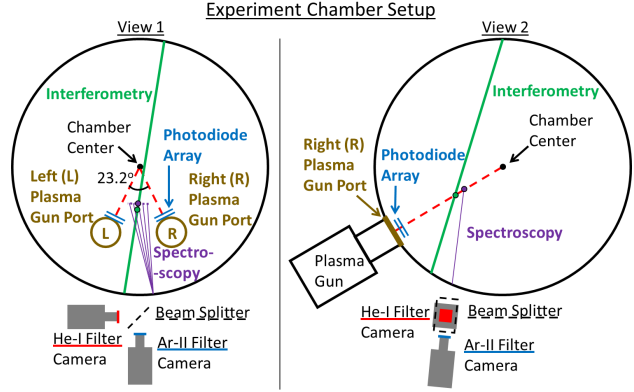}
	\caption{
		(Left) View 1: Projection of experimental setup, including approximate diagnostic views. Plasma guns mounted on ports of the spherical vacuum chamber launch plasma jets toward the chamber center in the direction shown by the red dashed lines. Due to expansion of the plasma jets, the jet collision takes place well before the chamber center, in the region observed by the interferometry and spectroscopy. The plasma guns are not shown in this View 1 for diagram clarity. (Right) View 2: Corresponding schematic with a 90$^{\circ}$ out-of-page rotation of View 1, adding a mounted plasma gun. Not to scale.
	}
	\label{fig:setup}
\end{figure}

We experimentally create multi-ion-species plasma shocks by colliding plasma jets generated from
plasma guns in a 2.7-m-diameter spherical vacuum
chamber \cite{Hsu2012, Merritt2013, Merritt2014, Moser2015, Hsu2018, Langendorf2018, Langendorf2019}. Figure ~\ref{fig:setup} depicts the experimental setup. Within the gun nozzle, a gas puff is pre-ionized into a plasma, and that plasma is accelerated by a 720 kA, 4.5~kV, 5~$\mu$s rise-time pulse. Initially, the gas puff atomic 
concentration, $c_{a}$, is 97\% He and 3\% Ar, corresponding to a mass concentration, $c_{m}$, of 76.4\% He 
and 23.6\% Ar. As the plasma jets propagate toward the chamber center, they expand in vacuum at 
approximately the sound speed. Well before reaching the chamber center, the two plasma jets merge at a half-angle of 11.6$^{\circ}$, with speeds of 55~$\pm$~5~km/s 
corresponding to 11~km/s in the direction normal to the shock and, therefore, a relative normal speed
between the jets of $v_{rel} = 22$~km/s. Before the shock formation, individual jets have electron number 
densities $n_{e} \approx 5\pm1\times10^{14}$~cm$^{-3}$, electron and ion temperatures $T_{e} \approx
T_{i} \approx 2.2^{+0.2}_{-0.5}$~eV, and inferred
average ionization states $\bar{Z}$ of $0.73^{+0.15}_{-0.67}$ for He and $1.15^{+0.23}_{-0.16}$ for Ar. These parameters give pre-shock Mach 
numbers $M=v_{rel}/[\gamma(\bar{Z}T_e+T_i)/m_i]^{1/2}$ 
of 1.7 for He ions, 5.1 for Ar ions, and an average $M= 1.8$, where we assume $\gamma = 5/3$ and a minimum ion $\bar{Z}=1$. The jets have characteristic scale 
lengths of $\sim$10~cm. Since the interspecies mean free path is much smaller than this overall characteristic length scale (see Sec.~\ref{sec:lengths}) and the thermal equilibration timescale is much smaller than the characteristic hydrodynamic timescale, the He and Ar should act as a single fluid with $T_{e} \approx T_{He} \approx T_{Ar} \approx T_{i}$ in the absence of gradients, i.e. pre-shock.

\begin{figure}[!tb]
	\includegraphics[width=3.0truein,keepaspectratio]{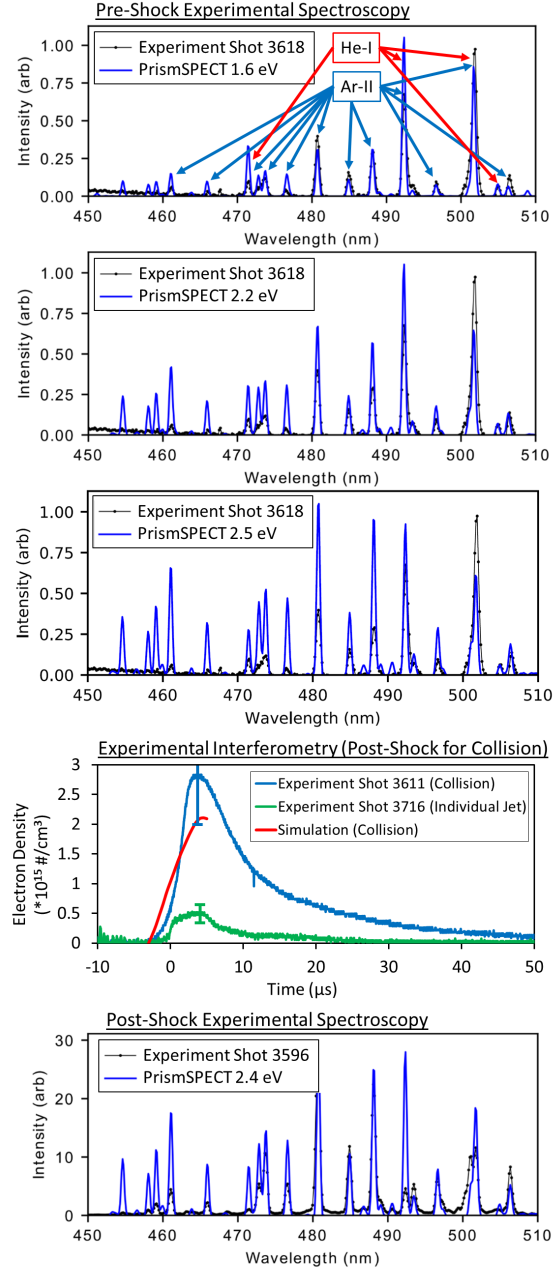}
	\caption{
		(Top 3 plots) Comparison of pre-shock experimental emission spectroscopy (black) with \textsc{prismspect} atomic-modeling calculations (blue). (Row 4) Comparison of experimental time-resolved interferometry electron-density measurements of one individual jet (green), which represents the pre-shock density for a collision, the central density (post-shock) of two jets merging (blue), and the central density (post-shock) in the simulation of two jets merging (red). Times are shifted such that $t=0$
		is the time at which the jets are first starting to merge. (Bottom) Comparison of post-shock experimental emission spectroscopy (black) with \textsc{prismspect} atomic-modeling calculations (blue).
	}
	\label{fig:diag}
\end{figure}

The plasma parameters are measured with diagnostics including a photodiode 
array to infer jet velocity via time-of-flight,
time-resolved interferometry for density \cite{Merritt2012, Merritt2012a}, and emission 
spectroscopy for temperature and ionization states. The photodiode array consists of two photodiodes per gun spaced 2~cm apart along the jet propagation direction, and jet speed is determined by looking at relative timing of the plasma emission intensity between the two photodiodes. The heterodyne interferometer measures time-resolved line-integrated electron density by observing relative phase shifts of a laser passing through the plasma compared to a reference path. We take the volumetric number density to be equal to the line-integrated areal density divided by the path length, and the path length is estimated based on the plasma emission length observed by visible camera images (not shown) taken along a direction perpendicular to the interferometry. The density measurement uncertainty is based upon these estimates of the path length. Spectroscopy data for the optically thin jets and generated shocks are compared with \textsc{prismspect} non-local-thermodynamic-equilibrium (non-LTE) atomic-modeling calculations \cite{PrismComputationalSciences}, as used in earlier research \cite{Hsu2012, Merritt2013, Merritt2014,Moser2015, Hsu2018, Langendorf2018, Langendorf2019}. Figure~\ref{fig:diag} displays the experimental temperature and density diagnostic results. As listed in Table~\ref{tab:param}, the pre-shock, experimentally inferred parameters are used to interpret the
experimental observations and as input for simulations.

\begin{table}[tb]
	\caption{Pre-shock and (simulation) post-shock parameters.}
	\label{tab:param}
	\begin{tabular}{lll} \hline\hline
		Parameter & Pre-shock & Post-shock \\ \hline
		Electron Temperature (eV) & 2.2 & 2.4 \\
		Electron Density ($\times 10^{14}$ cm$^{-3}$) & 5 & 21 \\ 
		He Temperature (eV) & 2.2 & 4.6 \\ 
		Ar Temperature (eV) & 2.2 & 5.5 \\ 
		He Average Ionization $\bar{Z}$ & 0.73 & 0.89 \\
		Ar Average Ionization $\bar{Z}$ & 1.15 & 1.47 \\
		He Atomic Concentration $c_{a1}$ & 97\% & 95.8\% \\
		Ar Atomic Concentration $c_{a2}$ & 3\% & 4.2\% \\
		He Mass Concentration $c_{m1}$ & 76.4\% & 69.6\% \\
		Ar Mass Concentration $c_{m2}$ & 23.6\% & 30.4\% \\
		\hline\hline
		
	\end{tabular}
\end{table}

\begin{figure}[!tb]
	\includegraphics[width=3.0truein,keepaspectratio]{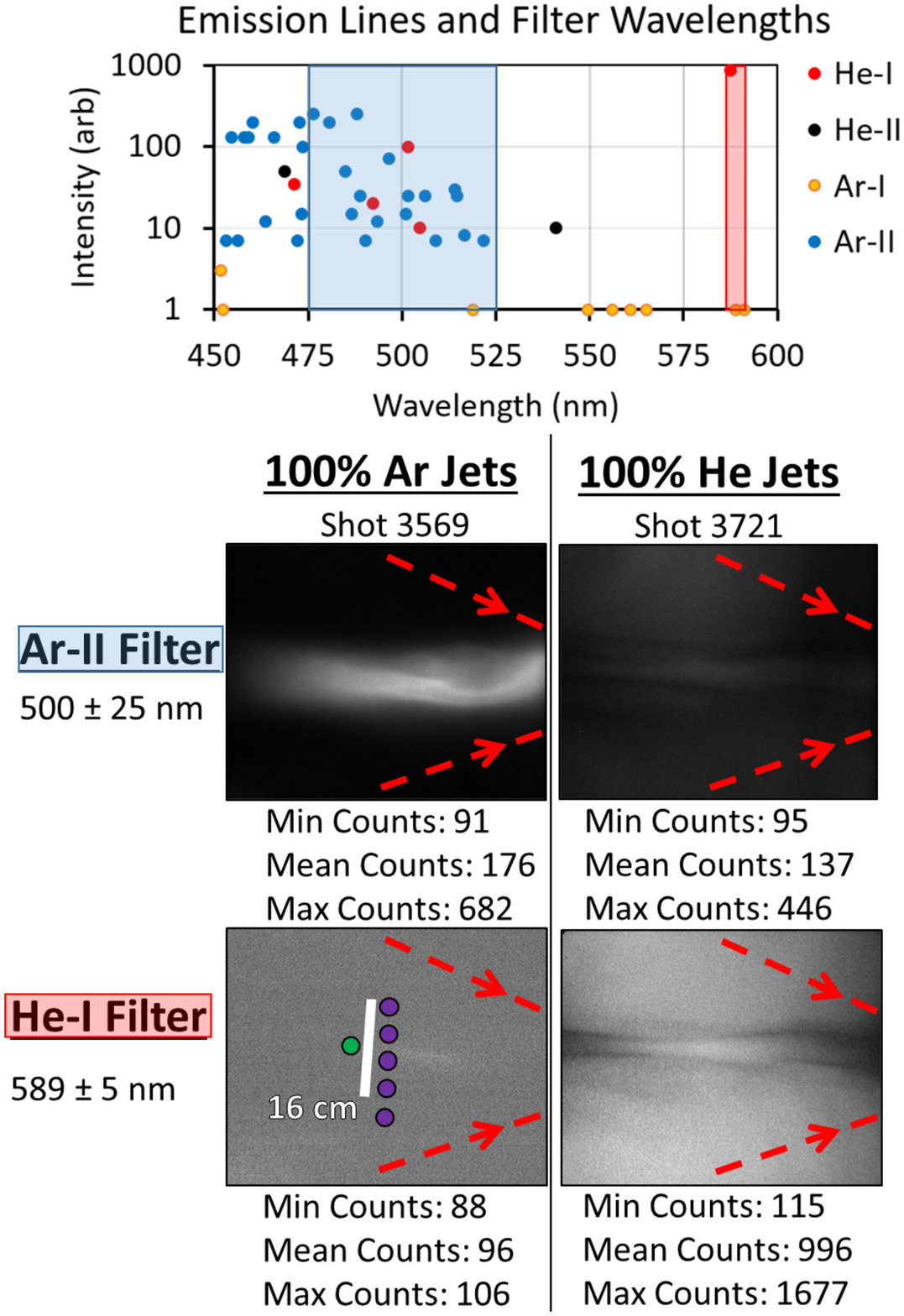}
	\caption{
		(Top) Plot of NIST strong emission lines (He-I, He-II, Ar-I, Ar-II) \cite{NIST} and the spectral filter wavelengths. (Bottom 2 Rows) To establish a baseline, time-gated images (10-ns exposure)
		of {\em single-ion-species} plasma jet merging
		using narrow bandpass filters are shown. The 100\%-Ar (left column) and
		100\%-He (right column)
		plasma shocks are much more prominently observed using the Ar-II and He-I filters, respectively. Jet merging and shock generation 
		occur near the centers of the images. Red dashed arrows show the direction of individual plasma jet propagation toward the chamber center (as in Fig.~\ref{fig:setup}). In the bottom left image, dots correspond to the intersection locations of the jet collision plane with the spectroscopy (purple) and interferometry (green) 
		lines of sight (as in Fig.~\ref{fig:setup}). The listed counts are over the whole image, for which 4096 counts is fully saturated.
	}
	\label{fig:filters}
\end{figure}

The shock profiles of the different ion species are imaged using a beam splitter to aim the plasma self-emission onto two intensified-charge-coupled-device (ICCD) cameras (10-ns exposure) with narrow bandpass 
filters \cite{Stenson2008, Stenson2012}. Singly ionized Ar-II line emission (near the
Ar-II 476.5--514.5-nm lines) is observed with a $500 \pm 25$-nm filter, and neutral He-I line emission 
(near the He-I 587.6-nm line) is observed with a $589 \pm 5$-nm filter \cite{NIST}. A filter for singly ionized He-II line 
emission was not used due to the better ICCD camera sensitivity to visible compared to ultraviolet 
wavelengths and the presence of other stronger lines (Ar-II and/or He-I) near the visible He-II lines. Figure~\ref{fig:filters} illustrates how the distinct filters are sensitive to
the different ion species, based on reference experiments using
single-ion-species plasma jet merging. For the 100\%-Ar jet merging, the bright region in the Ar-II filtered
image correlates to an increase in singly ionized Ar emission. For the 100\%-He jet collision, dark bands in the 
He-I filtered image correlate to a reduction in neutral He emission and a corresponding increase in He ionization. Thus, using these filters, we can separately image the shock profiles of the different ion 
species.

\section{\label{sec:exp}Experimental Shock Profile Results}

\begin{figure}[!]
	\includegraphics[width=3.25truein,keepaspectratio]{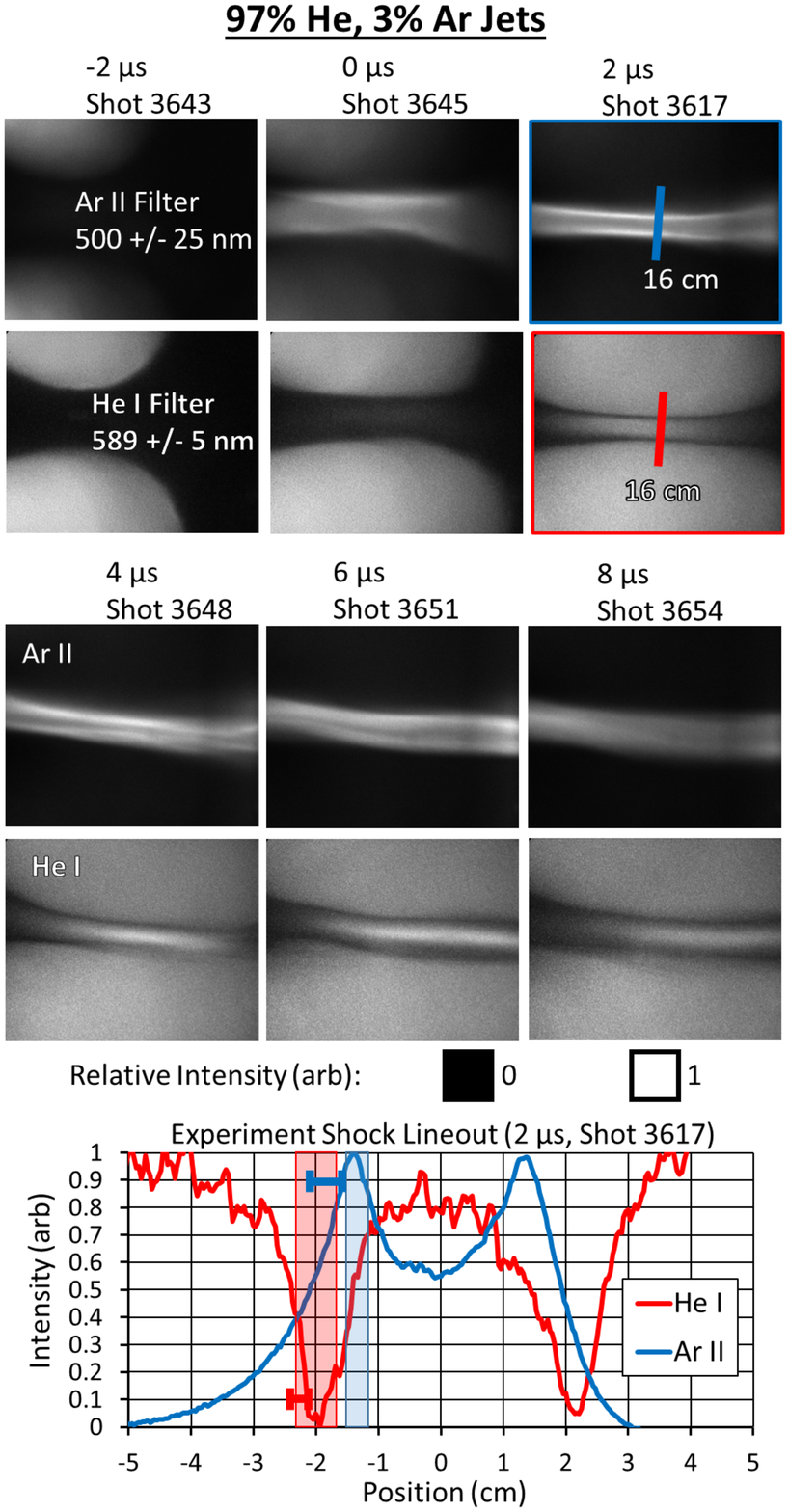}
	\caption{
		(Top 4 rows) Multi-ion-species plasma jet merging time evolution images of plasma self-emission using the Ar-II and He-I filters. Times defined such that $t=0$
		is the time at which the jets are first starting to merge. (Bottom row) Lineouts from the 2~$\mu$s images (in top rows) for the Ar-II 
		(blue line) and He-I (red line) filtered images; peak Ar-II intensity is at intensity of
		1, and peak He-II intensity is at intensity of 0. Horizontal bars show the shock widths: the distances between the 50\%--90\% intensity values. Shaded regions show the standard deviation of distance between peak ion intensities and the zero position.
	}
	\label{fig:multi_exp}
\end{figure}

By imaging the experimental shocks, we can infer spatial separation of the ions within the shock profiles. Our spatially resolved shock profiles also allow us to observe different shock widths for the different species.

The top 4 rows in Fig.~\ref{fig:multi_exp} show the time evolution of the merging of multi-ion-species plasma jets, observing the plasma self-emission using the Ar-II and He-I narrow bandpass filters. The time sequence is composed of images from different experimental shots, rather than images from multiple times during a single experimental shot. The jets originate from the top left and bottom left in the images, and they merge near the image center. The 0~$\mu$s time is defined as the time when we observe the jets first starting to merge and form a shock structure. Utilizing the interferometry data, we measure shot-to-shot timing variations of $<0.5~\mu$s. As the time sequence progresses, the individual jets continue to propagate at 11~km/s inward (toward image horizontal center) and $\sim$54~km/s parallel to the shock front (toward the image right). We use the intensity minima and maxima to infer the shock position. The plasma emission intensity is a function of electron temperature $T_{e}$ and ion density $n_{i}$, with larger temperatures and larger densities leading to higher ionization states and correspondingly more emission. We expect the maximum intensity of the shock to be located very near the locations of maximum temperature and density, but the maximum temperature may not be located precisely at the maximum density position. The shock generates sharp gradients in density and temperature that will create sharp gradients in intensity. Although the intensity gradient does not map one-to-one with each parameter across the shock, we examine the intensity gradient to indicate the shock gradient location and length scale.

Qualitatively in the top 4 rows in Fig.~\ref{fig:multi_exp}, we see the jets merging to form two intensity peaks originating a finite distance away from the image center. The experimental intensity peaks illustrate an increase in Ar-II and decrease in He-I (corresponding to an increase in He-II) within the shock. Then, near the center of the collision region, the amount of He-I increases and Ar-II decreases. This double-peaked shock profile has been observed in prior work studying single-ion-species plasma jet collisions with a similar plasma gun setup \cite{Merritt2013, Merritt2014}. Although this double-peaked feature is not as clearly seen in the shock created from the 100\% Ar jet collisions shown in Fig.~\ref{fig:filters}, the purpose of those images in Fig.~\ref{fig:filters} was to check the effectiveness of our spectral filters, so we did not spend time to experimentally tune the system in order to try to obtain better looking pictures. Compared to the 100\% He jet collision in Fig.~\ref{fig:filters}, the 97\% He and 3\% Ar multi-ion-species jet collision in Fig.~\ref{fig:multi_exp} shows more emission using the Ar-II filter.

The bottom row in Fig.~\ref{fig:multi_exp} displays the spatial profile of the shock structure produced by 
merging multi-ion-species plasma jets, which are directly imaged using the narrow bandpass filters for Ar-II 
and He-I line emission. Normal to the shock front, we take lineouts with widths of 20 pixels ($\sim$1~cm). In the 
laboratory frame, the shock fronts are moving away from the zero position. Across the emission spatial profile, 
the lineout intensities are scaled between 0 (lowest emission) and 1 (peak emission). For the 
Ar-II filtered lineout, 0 corresponds to background (lowest emission of singly ionized Ar), and 1 
corresponds to the peak singly ionized Ar emission. For the He-I filtered lineout,
1 corresponds to background 
(peak emission of neutral He), and 0 corresponds to peak singly ionized He emission (lowest 
emission of neutral He). Using these lineouts, we infer the shock-front separation between the peak 
intensities of Ar and He ions, showing that the He has moved farther ahead of the shock (closer to the pre-shock region) than the Ar. Additionally, we observe the shock width (horizontal bars in the bottom row of Fig.~\ref{fig:multi_exp} near the $-2$~cm position), which we approximate by the transition 
distance between the middle and peak intensity, here taken as 50\%--90\% of the peak value. 
The 50\% and 90\% 
values are chosen to reduce the influence of varying background levels and spurious pixel intensities on the 
measurements. These experimental results are obtained using the shock profile at 2~$\mu$s after the beginning of the 
plasma jet merging. This time is chosen by analyzing an image time sequence and 
correlating the density increase over time between the experimental time-resolved interferometry and the 
simulation results; see row 4 of Fig.~\ref{fig:diag}.

Using the shock profile lineouts and plasma parameters, we compare the length scales of our collisional plasma shocks. Taking statistics from 34 experimental shock-front profiles at the 2~$\mu$s time, we find a
separation between the shock-front intensity peaks to be $0.68 \pm 0.17$~cm, a He shock width of 
$0.36 \pm 0.09$~cm ($37 \pm 9$ post-shock He ion-ion mean free paths), and an Ar shock width of $0.52 \pm 
0.11$~cm ($57 \pm 12$ post-shock Ar ion-ion mean free paths). Additionally, we find the distance between the peak He ion intensity and the zero position to be $2.01 \pm 0.33$~cm, and the distance between the peak Ar ion intensity and the zero position to be $1.33 \pm 0.17$~cm; these distances are shown as the shaded regions in the bottom row of Fig.~\ref{fig:multi_exp}. Lastly, we find the ratio of the Ar shock width to He shock width to be $1.52 \pm 0.34$.  These experimental uncertainty values are the 
shot-to-shot standard deviations. The experimental shock widths are in reasonable agreement to the
theoretical prediction of $>$20 post-shock ion--ion mean free paths (for $M < 2$) \cite{Jaffrin1964, Amendt2011}. Furthermore, similar observations to our measurements of the ratios of shock widths for different species have been reported for neutral gas shocks, for which barodiffusion is present, and the lighter gas species also diffuses farther ahead of the neutral gas shock front compared to the heavier species \cite{Center1967}.

\section{\label{sec:lengths}Comparison of Shock Length Scales to Mean Free Path and Interpenetration}

Here, we compare the quantitative distances measured in the experiments to relevant length scales for the process of shock formation. This comparison allows us to better understand the physical picture.

Table~\ref{tab:lengthscales} compares the characteristic length scale results for the multi-ion-species 
plasma jet merging. Post-shock length scales are calculated based on the following plasma conditions obtained from simulations (Sec.~\ref{sec:sim}) and listed in Table~\ref{tab:param}: peak electron number density $n_{e} = \num{2.1e15}$~cm$^{-3}$, peak 
$T_{e} = 2.4$~eV, peak $T_{i,He} = 4.6$~eV, peak $T_{i,Ar} = 5.5$~eV, and 
average ionization states $\bar{Z}$ of 0.89 for He and 1.47 for Ar. The post-shock mean free path for an ion (unprimed)
colliding with multiple ion species (primed)
is \cite{NRL, Trubnikov1965}
\begin{equation}
L_{mfp,i} = \frac{v_{th,i}}{\sum_{i'} (\frac{n_{i'}}{n_{tot,i}}) \nu_{\epsilon}^{i,i'}} 
\label{eqn:mfp}
\end{equation} 
with thermal velocity $v_{th,i}$ and energy loss 
collision rate $\nu_{\epsilon}^{i,i'}$. Since we are not in the fast or slow limit, we use the full Coulomb collision-rate formula \cite{NRL}. In this formula, because we add contributions from individual ion species with discrete ionization states, $Z$, rather than using an average $\bar{Z}$, the collision rate is weighted by the individual species number density fraction $n_{i'}/n_{tot,i}$. We compare the experimentally
inferred shock width (as defined above in Sec.~\ref{sec:exp}) to the ion--ion mean free path rather than to the electron--ion
mean free path because the density jump and ion temperature jump within a shock front 
is predicted to occur over a distance of a few ion--ion mean free paths, while the electron temperature jump is predicted to occur over a (longer) distance of a few electron--ion mean free paths \cite{Jaffrin1964}. Furthermore, we calculate that the ion collision rates are dominant compared with the neutral collision rates $\nu_{n} \approx n_{n}\pi d_{n}^2 v_{th,n}$, where $d_{n} \sim 0.3$~nm is the approximate Van der Waals atomic diameter, and also with the neutral charge exchange collision rates $\nu_{CE} \approx n_{n} \sigma_{CE} v_{th,n}$, where $\sigma_{CE} \approx 3 \times 10^{-15}$~cm$^{2}$ is the charge exchange cross section \cite{Helm1977, Smirnov2000}.

\begin{table}[!tb]
	\caption{Plasma-jet-merging length scales (in centimeters).}
	\label{tab:lengthscales}
	\begin{tabular}{llll} \hline\hline
		Length scale (cm) & He & $\Delta$ & Ar \\ \hline
		He-Ar shock-front separation (exp.) & & 0.68 & \\ 
		He-Ar shock-front separation (sim.) & & 0.50 & \\
		Shock width (exp.)\ & 0.36 & & 0.52  \\ 
		Shock width (sim.)\ & 0.44 & & 0.57 \\ 
		Post-shock mean free path (Eq.~\ref{eqn:mfp}) & 0.0097 & & 0.0092 \\
		Jet characteristic size & 10 & & 10 \\
		Pre-shock interpenetration (Eq.~\ref{eqn:ionslowing}) & 0.058 & & 0.740  \\
		\hline\hline
		
	\end{tabular}
\end{table}

The ion--ion interpenetration length is the pre-shock slowing distance
over which ions from an individual jet (test particle,
unprimed) will stream through the other jet (field particles, primed) before becoming collisional
\cite{NRL, Trubnikov1965, Messer2013, Merritt2014, Moser2015, Langendorf2018, Langendorf2019},
\begin{equation}
L_{s,i} = \frac{v_{rel}}{4 \sum_{i'} (\frac{n_{i'}}{n_{tot,i}}) \nu_{s}^{i,i'}},
\label{eqn:ionslowing}
\end{equation} 
with relative velocity $v_{rel}$ and slowing-down collision rate $\nu_{s}^{i,i'}$. Again, we use the entire 
collision-rate formula \cite{NRL}. In our parameter space, the ion--ion slowing dominates
over ion--electron slowing, and thus the latter is ignored. Because our jet size is much larger than the interpenetration distance, the merging of the two supersonic plasma jets produces a shock.

\section{\label{sec:sim}Simulation Shock Profile Results}

We now compare the shock profile, length scales, and emission intensities obtained in the experiment to simulation results. The shock profile lengths in the simulations agree with the experimental observations within uncertainty bars. There are disagreements between the emission intensities in the post-processed simulations compared to the experiments, particularly in the post-shock region.

\subsection{\label{sec:sim}Comparison of Simulation and Experimental Shock Length Scales} 

We have
employed multi-fluid simulations, using the \textsc{chicago} code \cite{Thoma2011, Thoma2013, Thoma2017}, in which each plasma jet is modeled as
a separate ion fluid that can contain different species, to model the plasma collision \cite{Merritt2013,Langendorf2018,Langendorf2019}, including the interpenetration effects described in Sec.~\ref{sec:lengths}. \textsc{chicago} is a hybrid particle-in-cell code with the capability to model
both ions and electrons as fluid species \cite{Thoma2011}. The multi-fluid simulation can model interpenetration effects because each jet is treated as a separate fluid, and the jets can exchange momentum and energy via the inter-fluid collision models used in the code (see Eqs. 1 and 2 in Thoma et al., Physics of Plasmas 18, 103507 (2011) \cite{Thoma2011}). The code may also be run in a
magnetohydrodynamic (MHD) mode in which electron inertia is neglected.
In the present research, the multi-ion MHD approach was found to generate
identical results as those of lengthier simulations with electron inertia
retained. For the 1D simulation presented in this work, the two jets are given an initial density profile, temperature, and velocity based on input from experimental measurements. The simulation grid cell length is 0.06~cm, and the shock profile results are similar to results from a simulation with a 0.015~cm grid cell length. \textsc{propaceos} non-LTE data are used for the ion equations of state and opacities \cite{PrismComputationalSciences}. Collisionality between ions and electrons is determined by a Spitzer model. At initial temperatures below 2~eV, there was no discernible interaction between the colliding jets because the low ionization states lead to large interpenetration lengths. The simulation may need to be better tuned in these low ionization regimes and include neutral collision rates in order to obtain more physically accurate results.

\begin{figure}[!]
	\includegraphics[width=3.0truein,keepaspectratio]{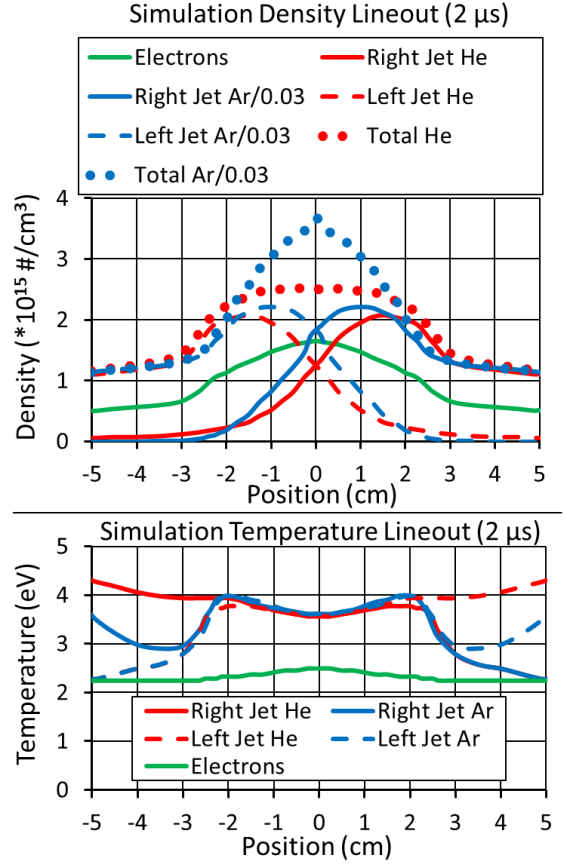}
	\caption{
		(Top) Simulated density lineouts for the
		He and Ar in jets coming from the left (dashed lines) and right (solid lines). Total densities (dotted lines) are the sum of the left and right jets. Ar density values are
		divided by 0.03 for visualization. (Bottom) Corresponding simulated temperature lineouts. Times are at 2~$\mu$s after jets first 
		start to merge.
	}
	\label{fig:multi_sim}
\end{figure}

The \textsc{chicago} 1D multi-fluid simulations agree with the experimental results, showing: shock-front separation, He diffusing farther ahead of the shock (closer to the pre-shock region) than Ar, and different shock widths for different ions. Figure~\ref{fig:multi_sim} displays the simulated densities and temperatures, respectively,
for He and Ar in the individual plasma 
jets coming from the left and right (dashed and solid lines) at 2~$\mu$s after the jets start to merge. We use these individual jet shock profiles to obtain the following quantities. Simulations indicate a separation 
between the shock-front density peaks to be $0.50 \pm 0.12$~cm, a He shock width of $0.44 \pm 0.12$~cm 
($45 \pm 12$ post-shock He ion-ion mean free paths), and an Ar shock width of $0.57 \pm 0.12$~cm ($62 \pm 13$ post-shock Ar 
ion-ion mean free paths). Additionally, we find the distance between the peak He density and the zero position to be $1.51 \pm 0.06$~cm, and the distance between the peak Ar density and the zero position to be $1.01 \pm 0.06$~cm. We also find the ratio of the Ar shock width to He shock width to be $1.29 \pm 0.45$. The simulation uncertainty values are based upon the 0.06~cm simulation resolution. Except for the distances between the peak intensities and the zero position, these simulation quantities agree with the experimental data within uncertainty bars.

\subsection{\label{sec:synth}Comparison of Simulation and Experimental Intensity Profiles}

Through this point in the paper, the experiments and simulations have generally agreed. Because the experimental results in Fig.~\ref{fig:multi_exp} are emission intensity and the simulation results in Fig.~\ref{fig:multi_sim} are density and temperature, it is useful to directly compare the experiments with synthetic emission intensities from the simulation. However, in the simulation post-shock region, there are significant disagreements to the experiment regarding the plasma state: particularly the ionization state and emission intensity. The simulation post-shock density prediction matches relatively well based on the experimental interferometry (see Fig.~\ref{fig:diag}). The simulation post-shock ion temperature predictions are reasonable based on calculations converting 100\% of the ion kinetic energy (at 11~km/s) into thermal energy within the shock, which would generate post-shock $T_{He_{max}} \approx 4.7$~eV and $T_{Ar_{max}} \approx 27.5$~eV, and we expect post-shock temperatures closer to the $T_{He_{max}}$ value due to the relative atomic concentrations of the species. Post-shock ion temperatures were also measured in experiments for single ion species shocks, showing agreement with simulation results \cite{Langendorf2018, Langendorf2019}. Although the simulation post-shock density and temperature values, which influence the intensity results, seem reasonable, the following paragraph will explore differences in the simulation ionization states and synthetic emission intensities compared to the experiments.

\begin{figure}[!]
	\includegraphics[width=3.0truein,keepaspectratio]{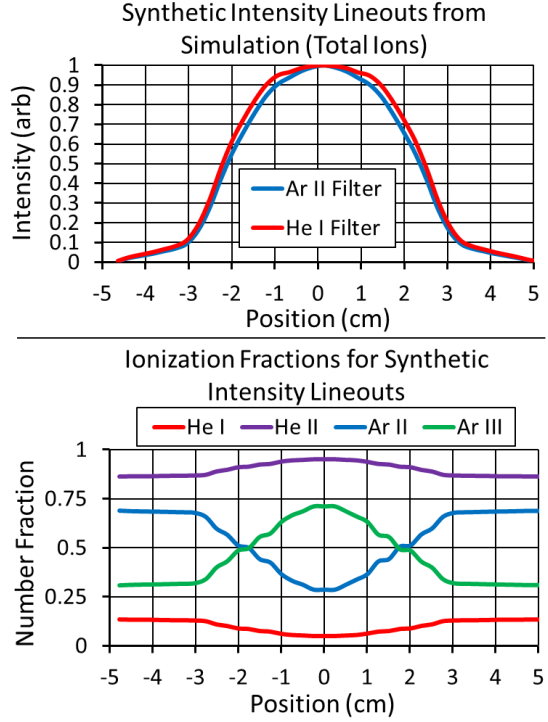}
	\caption{
		(Top) Synthetic intensity lineouts of plasma self-emission using the Ar-II ($500 \pm 25$~nm) and He-I ($589 \pm 5$~nm) filters with \textsc{spect3D} \cite{PrismComputationalSciences}. Inputs for these intensities are the simulation total densities and temperatures shown in Fig.~\ref{fig:multi_sim}. Comparing with Fig.~\ref{fig:multi_exp}, the synthetic lineouts do not depict the experimentally observed double-peaked intensity feature. (Bottom) The ionization fractions of neutral He (He I), singly ionized He (He II), singly ionized Ar (Ar II), and doubly ionized Ar (Ar III) corresponding to the above synthetic intensity lineouts. The average ionization values for both He and Ar increase near the center post-shock region, despite the He-I filter intensity remaining large.
	}
	\label{fig:synth}
\end{figure}

While the experiments and simulations quantitatively agree within uncertainty bars regarding the magnitude of the shock-front separation and the He and Ar shock widths, the simulations predict a single-peaked shock structure near the collision center rather than the double-peaked and wider post-shock region observed in the experiments; see the dotted lines in the top row of Fig.~\ref{fig:multi_sim} compared to the bottom row of Fig.~\ref{fig:multi_exp}. This difference correlates with the smaller distance between the center zero position and the ion density peaks found in the simulation compared to the experiments. As shown in the top row of Fig.~\ref{fig:synth}, we use \textsc{spect3D} software \cite{PrismComputationalSciences} to obtain synthetic intensity lineouts of the \textsc{chicago} simulation density and temperature parameters. This synthetic intensity lineout does not reproduce the experimentally observed double-peaked shock profile or the decreases in relative intensity near the zero position for either Ar-II or He-I filter. Still, as shown in the bottom row of Fig.~\ref{fig:synth}, the average ionization values for both species are increasing in the post-shock region near the center position of the plasma jet merging, as is expected. A 2D Cartesian simulation of the jet merging process provides qualitatively similar results as the 1D simulation, also depicting a single-peaked shock structure.

These differences in the post-shock region between experiments and simulation will be further assessed in future work. The disagreement seems to occur from the simulation predicting larger amounts of He-I (rather than He-II) in the shock compared to the experiment and also not showing a region of decreased Ar-II near the center of the collision region. Possible sources of this discrepancy include the models of collisionality and thermal conduction in the post-shock region used in the simulations for this plasma parameter regime. Additionally, the simulation EOS and opacity models do not capture time-dependent effects of ionization and/or recombination during the transient process of shock formation, as is present in the experiments. Ultimately, although the experimental and simulated emission intensities disagree in the post-shock region, the separation of ion species within the shock front agrees well between the simulations and experimental observations.

\section{\label{sec:diff}Dominance of Barodiffusion in the Present Work}

The shock-front separation results are consistent among
the experiments, simulations, and theoretical predictions of ion species diffusion. In this section, we quantitatively estimate the expected species concentration change within the shock and calculate the relative contributions of the different diffusion mechanisms, showing that barodiffusion dominates in the present research. 

As stated earlier in Sec.~\ref{sec:background}: within a shock front, theory predicts that the lighter ion diffusion velocity (in the center of mass frame) points toward the pre-shock region, and the heavier ion diffusion velocity points in the opposite direction. This prediction is consistent with our experimental and simulation results for the He/Ar mixture. The interspecies ion diffusion should change the relative species concentration. Experimentally, we did not directly measure the species concentration along the shock profile, but the simulations show interspecies ion diffusion. Compared to the initial condition of 97\% He and 3\% Ar atomic concentrations [76.4\% He and 23.6\% Ar mass concentrations], in the top row in Fig.~\ref{fig:multi_sim} we find a minimum He atomic concentration in the post-shock region (corresponding to the zero position) of 95.8\% [69.6\% He mass concentration], and we find a maximum He atomic concentration at the shock front closer to the pre-shock region (corresponding to the $\pm~2.5$~cm position) of 97.7\% [80.7\% He mass concentration].

We now quantitatively explore the effects of different diffusion flux terms on the species concentration. The diffusion terms depend upon the species concentration, masses, number densities, ionization states, and temperatures. The pre-shock and post-shock plasma parameters are listed in Table~\ref{tab:param}. The goal of the following calculations are to serve as order-of-magnitude estimates to show consistency between the theory, experiments, and simulations. First, assuming no initial concentration separation for simplicity, we take $\nabla c_{m1} = 0$ in Eq.~\ref{eqn:conc} and Eq.~\ref{eqn:diffusion}. We also approximate $\partial t = 2~\mu$s. Taking a characteristic length scale as the shock width $L_{SW} \approx 0.4$~cm, we estimate $\nabla = 1/L_{SW}$. Therefore, for a quantity Q in Eq.~\ref{eqn:diffusion}, we have $(\nabla Q) / Q \approx (1/L_{SW}) (Q_{post-shock} - Q_{pre-shock}) / Q_{pre-shock} $. For our He/Ar mixture, we then calculate the diffusion coefficients to be D = 600~cm$^2$/s, $\kappa_{p} = 1.3$, $\kappa_{Ti} = 0.5$, $\kappa_{Te} \approx -1$, and $\kappa_{e} = 1.3$ \cite{Kagan2012, Simakov2017, Kagan2019, Zhdanov2002}. See Appendix~\ref{sec:app_diff} for expressions for the diffusion coefficients. We also estimate the ambipolar electric field $\nabla \phi \approx \nabla T_{e} / e$ \cite{Amendt2009,Tang2014}. Therefore, the relative contributions for each diffusion mechanism (the quantities in parentheses in Eq.~\ref{eqn:diffusion}) are 20.3~cm$^{-1}$ for barodiffusion, 1.14~cm$^{-1}$ for thermodiffusion (1.36~cm$^{-1}$ for ion thermodiffusion and -0.23~cm$^{-1}$ for electron thermodiffusion), and 0.30~cm$^{-1}$ for electrodiffusion. Starting with a He concentration $c_{m1} = \rho_{1}/ \rho = 76.4\%$, in Eq.~\ref{eqn:flux} we obtain a He diffusion speed $v_{D1} = 1.7 \times 10^{-2}$~cm/$\mu$s, corresponding in Eq.~\ref{eqn:conc} to a change in concentration of $\partial c_{m1} = 6.5\%$. This estimated calculation result is in agreement with the simulation concentration change from the above paragraph. Furthermore, these estimates show that barodiffusion is the dominant diffusion mechanism in the present research that causes the shock-front separation. This result contrasts with prior ICF experiments, for which the ion thermodiffusion mechanism dominated \cite{Hsu2016, Joshi2017}. Additional experimental testing and validation of the theory can be achieved by changing the species concentrations, which influence the contributions of the various diffusion mechanisms.

\section{\label{sec:concl}Conclusion}

Within a collisional multi-ion-species
plasma shock front containing 97\% He and 3\% Ar, we experimentally observe 
shock-front separation and species-dependent shock widths. Experiments, 1D multi-fluid plasma
simulations, and theoretical predictions
are all consistent in showing that the lighter He ions diffuse farther ahead within the overall shock front than the heavier Ar ions. The experimental shock profiles of different ion species were directly imaged using narrow bandpass visible wavelength filters. Multi-fluid plasma simulations allowed for reasonably
accurate modeling of the 
plasma jet merging and multi-ion-species effects, though with disagreement on the emission intensities in the post-shock plasma. The fundamental experimental data in the present work can be used to validate models and benchmark numerical simulations of multi-ion-species
collisional plasma shocks of relevance to HED, MIF, and ICF experiments. These experimental data are a valuable start to the process of benchmarking codes in this regime by demonstrating where there is agreement and disagreement. Additional work can be performed to obtain species concentration measurements at multiple points across the shock profile during a single experimental shot. These measurements will likely require a more detailed spectroscopic study of the emission line ratios in comparison with the \textsc{prismspect} atomic modeling calculations. More experiments can be conducted in order to acquire data for other times during the shock propagation, which can provide information about time-dependent effects during the process of shock formation including ionization, recombination, and kinetic (non-Maxwellian) effects.

\begin{acknowledgments}
We acknowledge 
J. Dunn, K. C. Yates, S. Brockington, A. Case, E. Cruz, F. D. Witherspoon, Y. C. F. Thio, D. Welch, P. Bellan, G. Kagan, and X.-Z. Tang
for technical support and/or useful discussions. This work was supported by the Office of Fusion Energy Sciences and the Advanced Research Projects Agency--Energy of the U.S. Department of Energy under contract number DE-AC52-06NA25396.	
\end{acknowledgments}

\appendix
\section{\label{sec:app_diff}Diffusion Coefficient Calculations}
For reference, we list the equations to determine the diffusion coefficients \cite{Kagan2012, Simakov2017, Kagan2019, Zhdanov2002}, where subscript ``1'' refers to the lighter ion species with mass $m_{1}$, and subscript ``2'' refers to the heavier ion species with mass $m_{2}$. The classical diffusion coefficient, D, is
\begin{equation}
D = \frac{\rho T_{i}}{A_{12}\mu_{12}n_{1}\nu_{12}} \times \frac{c_{m1}(1-c_{m1})}{c_{m1}m_{2} + (1-c_{m1})m_{1}}
\label{eqn:D}
\end{equation}
where the new variables are the transport coefficient $A_{12}$ ($\sim 1$, here) and reduced mass $\mu_{12} = m_{1}m_{2}/(m_{1}+m_{2})$. The collision frequency $\nu_{12}$ is the energy loss collision rate. Next, the barodiffusion ratio, $\kappa_{p}$, is
\begin{equation}
\kappa_{p} = c_{m1}(1-c_{m1})(m_{2}-m_{1})(\frac{c_{m1}}{m_{1}} + \frac{1-c_{m1}}{m_{2}})
\label{eqn:kappa_p}
\end{equation}
The electrodiffusion ratio, $\kappa_{e}$, is
\begin{equation}
\kappa_{e} = m_{1}m_{2}c_{m1}(1-c_{m1})(\frac{c_{m1}}{m_{1}} + \frac{1-c_{m1}}{m_{2}})(\frac{Z_{1}}{m_{1}} - \frac{Z_{2}}{m_{2}})
\label{eqn:kappa_e}
\end{equation}
The ion thermodiffusion ratio, $\kappa_{Ti}$, is
\begin{equation}
\kappa_{Ti} = m_{1}m_{2}(\frac{c_{m1}}{m_{1}} + \frac{1-c_{m1}}{m_{2}})[\frac{c_{m1}B_{11}}{m_{1}} + \frac{(1-c_{m1})B_{12}}{m_{2}}]
\label{eqn:kappa_Ti}
\end{equation}
and the electron thermodiffusion ratio, $\kappa_{Te}$, is
\begin{equation}
\begin{split}
\kappa_{Te} = & m_{1}m_{2}(\frac{c_{m1}}{m_{1}} + \frac{1-c_{m1}}{m_{2}})[\frac{c_{m1}Z_{1}}{m_{1}} + \frac{(1-c_{m1})Z_{2}}{m_{2}}] \\ &\times [(1-c_{m1})B_{1e} - c_{m1}B_{2e}]\frac{T_{e}}{T_{i}}
\label{eqn:kappa_Te1}
\end{split}
\end{equation}
where $B_{11}$, $B_{12}$, $B_{1e}$, and $B_{2e}$ are transport coefficients that can be evaluated numerically. The electron thermodiffusion ratio, $\kappa_{Te}$, can also be written as
\begin{equation}
\begin{split}
\kappa_{Te} = & -Z_{1}^{2}c_{m1}(1-c_{m1})(c_{m1} + \frac{(1-c_{m1})m_{1}}{m_{2}})  \\ &\times (\frac{m_{2}}{m_{1}} - \frac{Z_{2}^{2}}{Z_{1}^{2}}) \frac{T_{e}}{T_{i}} \frac{\beta_{0}}{Z_{eff}}
\label{eqn:kappa_Te2}
\end{split}
\end{equation}
where the coefficient $\beta_{0}$ is
\begin{equation}
\beta_{0} = \frac{30 Z_{eff}(11 Z_{eff} + 15 \sqrt{2})}{217 Z_{eff}^{2} + 604 \sqrt{2} Z_{eff} + 288}
\label{eqn:beta}
\end{equation}
where $Z_{eff}$ is
\begin{equation}
Z_{eff} = \frac{n_{1}Z_{1}^{2} + n_{2}Z_{2}^{2}}{n_{1}Z_{1} + n_{2}Z_{2}}
\label{eqn:Zeff}
\end{equation}

\bibliography{arxiv_multispecies_PoP_byvank_final}

\end{document}